\documentstyle[aps,prl,multicol,epsf]{revtex}

\begin{document}   
\draft  
   
\title {Generic Dynamic Scaling in Kinetic Roughening}

\author{Jos\'e J. Ramasco$^{1,2,}$\cite{mail1}, 
Juan M. L\'opez$^{3,}$\cite{mail2} and
Miguel A. Rodr\'{\i}guez$^{1}$ 
}
 
\address{$^1$ Instituto de F\'{\i}sica de Cantabria, 
CSIC-UC, E-39005 Santander, Spain}  

\address{$^2$ Departamento de F\'{\i}sica Moderna,   
Universidad de Cantabria, E-39005 Santander, Spain}   

\address{$^3$ Dipartamento di Fisica and Unit\`a INFM,
Universit\`a di Roma "La Sapienza", I-00185 Roma, Italy}

%\address{$^4$ Department of Mathematics,   
%Imperial College, London SW7 2BZ, United Kingdom}                       
   
\maketitle    
    
\begin{abstract}   
We study the dynamic scaling hypothesis 
in invariant surface growth. We show that the existence of power-law 
scaling of the correlation functions (scale invariance)
does not determine a unique dynamic scaling  
form of the correlation functions, which leads to 
the different anomalous forms of scaling recently observed
in growth models.
We derive all the existing forms of anomalous dynamic 
scaling from a new generic scaling ansatz.
The different scaling forms are subclasses
of this generic scaling ansatz associated with bounds
on the roughness exponent values. The existence of 
a new class of anomalous dynamic scaling is predicted
and compared with simulations.
\end{abstract}   

%\pacs{05.40.+j,05.70.Ln,68.35.Fx} %Pacs already revised   

\begin{multicols}{2}
\narrowtext     
       
The theory of kinetic roughening 
deals with the fate of surfaces growing in
nonequilibrium conditions \cite{barabasi,krug-rev}. 
In a typical situation an initially flat surface 
grows and roughens continuously as it  
is driven by some external noise.
The noise term can be of thermal origin 
(like for instance fluctuations in the flux 
of particles in a deposition process), or a
quenched disorder (like in the motion of 
driven interfaces through porous media). 
A rough surface may be characterized by
the fluctuations of the height around
its mean value. So, a basic quantity to look at 
is the {\em global} interface width, 
$W(L,t) = \langle [\overline{h(x,t) - \overline h]^2}\rangle^{1/2}$,
where the overbar denotes 
average over all $x$ in a system of size $L$ and brackets 
denote average over different realizations. 
Rough surfaces then correspond to situations in which 
the stationary width $W(L,t \to \infty)$ grows with the system size. 
Alternatively, one may calculate other quantities 
related to correlations over a distance $l$ as 
the height-height correlation function, 
$G(l,t) = \langle \overline{[h(x+l,t) - h(x,t)]^2}\rangle$,
or the {\em local} width, 
$w(l,t) = \langle \langle [h(x,t) - 
\langle h \rangle_l]^2\rangle_l\rangle^{1/2}$, where
$\langle \cdots \rangle_{l}$ denotes an average over $x$ 
in windows of size $l$.  

In absence of any characteristic length in the
problem growth processes are expected
to show power-law behaviour 
of the correlation functions in space and time
and the Family-Vicsek dynamic
scaling ansataz \cite{family,barabasi,krug-rev}
\begin{equation}
\label{FV-globalwidth}
W(L,t) = t^{\alpha/z} f(L / \xi(t)),
\end{equation}
ought to hold. 
The scaling function $f(u)$ behaves as
\begin{equation}
\label{FV-forf}
f(u) \sim
\left\{ \begin{array}{lcl}
     u^{\alpha}     & {\rm if} & u \ll 1 \\
     {\rm const.} & {\rm if} & u \gg 1
\end{array}
\right.,
\end{equation}
where $\alpha$ is the roughness exponent and 
characterizes the stationary
regime, in which the horizontal
correlation length $\xi(t) \sim t^{1/z}$ 
($z$ is the so called dynamic exponent) has reached a value 
larger than the system size $L$. 
The ratio $\beta= \alpha/z$ is called growth exponent and
characterizes the short time behavior of the surface. 
As occurs in equilibrium critical phenomena, 
the corresponding critical exponents 
do not depend on microscopic details of the system
under investigation. This has made 
possible to divide growth processes
into universality classes according to the 
values of these characteristic exponents \cite{barabasi,krug-rev}.

A most intringuing feature of some growth models is that
the above standard scaling of the global width differs 
substancially from the scaling behaviour of the local 
interface fluctuations (measured either by the local
width or the height-height correlation). 
More precisely, in some growth models
the local width (and the height-height correlation) 
scales as in Eq.(\ref{FV-globalwidth}), 
{\it i.e.} $w(l,t)=t^\beta f_A(l/\xi(t))$,
but with the anomalous scaling function
\begin{equation}
\label{f_A}
f_A(u) \sim
\left\{ \begin{array}{lcl}
     u^{\alpha_{loc}}     & {\rm if} & u \ll 1 \\
     {\rm const} & {\rm if} & u \gg 1 
\end{array}
\right., 
\end{equation}
where the new independent exponent 
$\alpha_{loc}$ is called the local roughness exponent. 
This is what has been called {\em anomalous} roughening
in the literature, and has been found to occur 
in many growth models 
\cite{krug,schro,lopez96,das1,das2,lopez97c,mario} 
as well as experiments 
\cite{yang,jef,fracture1,fracture2,bru}.
Moreover ,it has recently been 
shown \cite{lopez97a,lopez97b} that anomalous 
roughening can take two different forms. 
On the one hand, there 
are {\em super-rough} processes,
{\it i.e.} $\alpha > 1$, for which always $\alpha_{loc} = 1$.
On the other hand, there are {\em intrinsically} anomalous 
roughened surfaces, for which $\alpha_{loc} < 1$ and 
$\alpha$ can actually be any $\alpha > \alpha_{loc}$.

Anomalous scaling 
implies that one more independent
exponent, $\alpha_{loc}$, 
may be needed in order to asses the
universality class of the particular system under study. In other
words, some growth models may have exactly the same $\alpha$
and $z$ values seemingly indicating that they belong to the same
universality class. However, they may have different values of
$\alpha_{loc}$ showing that they actually
belong to distict classes of growth. As for the experiments,
only the local roughness exponent is measurable by direct methods,
since the system size remains normally fixed. Fracture 
experiments \cite{fracture2} 
in systems of varying sizes have succeded in measuring both the
local and global roughness exponents in good agreement with the
scaling picture described above.

In this Letter we introduce a new
anomalous dynamics in kinetic roughening.
We show that, by adopting more general
forms of the scaling functions involved, 
a generic theory of dynamic scaling
can be constructed.
Our theory incorporates all the different forms
that dynamic scaling can take, namely Family-Vicsek,
super-rough and intrinsic, as subclasses and predicts
the existence of a new class of 
growth models with novel anomalous scaling properties.
Simulations of the Sneppen model (rule A) \cite{sneppen}
of self-organized depinning (and other related models) 
are presented as examples of the new dynamics.

Firstly, let us consider the Fourier transform of the 
height of the surface in a system of size $L$,
which is given by
$\widehat{h}(k,t) = 
L^{-1/2} \sum_x [h(x,t) - \overline h(t)] \exp(ikx)$,
where the spatial average of the height has been 
substracted. The scaling 
behaviour of the surface can now be investigated by 
calculating the structure factor or power spectrum 
\begin{equation}  
\label{S}
S(k,t) = 
\langle \widehat{h}(k,t) \widehat{h}(-k,t) \rangle,
\end{equation}
which is related to the height-height
correlation function $G(l,t)$ defined above 
by 
\begin{eqnarray}
G(l,t) & = & {4 \over L} \sum_{2\pi/L\leq k \leq \pi/a} [1-\cos (kl)] S(k,t)
\nonumber
\\
& \propto & \int_{2\pi /L}^{\pi/a} {dk \over 2\pi} [1 - \cos(kl)] S(k,t),
\label{G-from-S}
\end{eqnarray}
where $a$ is the lattice spacing and $L$ is the system size.

In order to explore the most general 
form that kinetic roughening can take,
we study the scaling behaviour of surfaces
satisfying what we will call
a {\em generic} dynamic scaling form of the
correlation functions.
We will consider that a growing surface satisfies
a generic dynamic scaling when there 
exists a correlation lenght 
$\xi(t)$, {\it i.e.} the distance over which 
correlations have propagated up to
time $t$, and 
$\xi(t) \sim t^{1/z}$, being $z$ the dynamic
exponent.
If no characteristic scale exists but
$\xi$ and the system size $L$, then 
power-law behaviour in space and time 
is expected and the growth
saturates when $\xi \sim L$ and the correlations
(and from Eq.(\ref{G-from-S}) also the structure factor)
become time-independent. The global roughness
exponent $\alpha$ can now be calculated 
in this regime from $G(l=L,t \gg L^z) \sim L^{2\alpha}$
(or $W(L,t \gg L^z) \sim L^\alpha$).
In general, as we will see below, the scaling 
function that enters the dynamic scaling of
the local width (or the height-height correlation)
takes different forms depending
on further restrictions and/or bounds for the
roughness exponent values. These kind of 
restrictions are very often assumed and not valid for
every growth model. 
For instance, only if the surface were 
{\em self-affine} saturation of the 
correlation function $G(l,t)$ would also occur 
for intermediate scales $l$ at times $t \sim l^z$
and with the very same roughness exponent.
However, the latter does not hold when anomalous
roughening takes place as can be seen from
the scaling of the local width in Eq.(\ref{f_A}). 

Our aim here is to investigate {\em all}
the possible forms that the scaling functions
can exhibit when solely the existence of generic
scaling is assumed. So, if the roughening 
process under consideration shows generic
dynamic scaling (in the sense above explained),
and no further assumptions (like for instance 
surface self-affinity or implicit bounds for the
exponents values) 
are imposed, then we propose that the 
structure factor is given by
\begin{equation}
\label{Skt}
S(k,t) =k^{-(2\alpha+1)} s(kt^{1/z}),
\end{equation}
where the scaling function has the general form
\begin{equation}
\label{Anom-s}
s(u) \sim  
\left\{ \begin{array}{lcl}
     u^{2(\alpha-\alpha_s}) & {\rm if} &  u \gg 1\\
     u^{2\alpha + 1} & {\rm if} &  u \ll 1
     \end{array}
\right. ,
\end{equation}
and the exponent $\alpha_s$ is what we will call the
{\em spectral} roughness exponent.
This scaling ansatz is a natural generalization of
the scaling proposed for the structure factor 
in Ref.\cite{lopez97a,lopez97b} for anomalous scaling. 

In the case of the
global width, one can make use of
\begin{equation}
W^2(L,t) = {1 \over L} \sum_k S(k,t) 
= \int {dk \over 2\pi} S(k,t),
\end{equation}
to prove easily that the global width scales as 
in Eqs.(\ref{FV-globalwidth}) and (\ref{FV-forf}),
independently of the value of the exponents $\alpha$
and $\alpha_s$.

However, the scaling of the local width is much more
involved. The existence of 
a generic scaling behaviour like (\ref{Anom-s})
for the structure factor always leads to a 
dynamic scaling behaviour, 
\begin{equation}
w(l,t) \sim \sqrt{G(l,t)}=t^\beta g(l/\xi) 
\end{equation}
of the height-height correlation (and local width), 
but the corresponding scaling function $g(u)$ 
is not unique. 
When substituting Eqs.(\ref{Anom-s}) 
and (\ref{Skt}) into (\ref{G-from-S}),
one can see that the various limits involved 
($a \to 0$, $\xi(t)/L \to \infty$ and 
$L \to \infty$) do not conmute \cite{lopez97a,lopez97b}.
This results in a different scaling behaviour of $g(u)$
depending on the value of the exponent $\alpha_s$. 

Let us now summarize how all
scaling behaviours reported in the 
literature are obained from the 
generic dynamic scaling ansatz (\ref{Anom-s}).
We shall also show how  
a new roughening dynamics naturally appears in this
scaling theory. Two major cases can be distinguised,
namely $\alpha_s < 1$ and $\alpha_s > 1$. 
On the one hand,
for $\alpha_s < 1$ the integral in 
Eq.(\ref{G-from-S}) has already been computed 
\cite{lopez97a,lopez97b} and one gets 
\begin{equation}
\label{a<1}
g_{\alpha_s<1}(u) \sim
\left\{ \begin{array}{lcl}
     u^{\alpha_{s}}     & {\rm if} & u \ll 1 \\
     {\rm const} & {\rm if} & u \gg 1 
\end{array}
\right..
\end{equation} 
So, the corresponding scaling function is 
$g_{\alpha_s<1} \sim f_A$ 
and $\alpha_s=\alpha_{loc}$, 
{\it i.e.} the intrinsic 
anomalous scaling function in Eq.(\ref{f_A}). 
Moreover,
in this case the interface would satisfy a
Family-Vicsek scaling (for the local as well 
as the global width) 
only if 
$\alpha=\alpha_s$ were satisfied for the
particular growth model under study.
Thus, the standard Family-Vicsek scaling turns 
out to be one of the possible scaling forms
compatible with generic scaling invariant growth,
but not the only one.

On the other hand, a new anomalous 
dynamics shows up for growth models in which 
$\alpha_s > 1$.
%In this case, one finds
In this case, one finds that, 
in the thermodynamic limit $L \to \infty$, 
the integral Eq.(\ref{G-from-S})
has a divergence coming from the lower 
integration limit. 
To avoid the divergence
one has to compute the integral keeping $L$ fixed.
We then obtain the scaling function
\begin{equation}
\label{a>1}
g_{\alpha_s>1}(u) \sim
\left\{ \begin{array}{lcl}
     u     & {\rm if} & u \ll 1 \\
     {\rm const} & {\rm if} & u \gg 1 
\end{array}
\right..
\end{equation} 
So that in this case one always gets
$\alpha_{loc}=1$ for any $\alpha_s > 1$.
Thus, for growth models in 
which $\alpha = \alpha_s$ 
one recovers the super-rough 
scaling behaviour \cite{lopez97a,lopez97b}.

However, it is worth noting that 
neither the spectral exponent $\alpha_s$ 
or the global exponent $\alpha$ are fixed 
by the scaling in Eqs(\ref{Anom-s}) 
and (\ref{a>1})
and, in principle, they could be different. 
Therefore, growth models in which 
$\alpha_s > 1$ but $\alpha \ne \alpha_s$ could also
be possible and represent a new type of dynamics
with anomalous scaling. 
The main feature of this new type of anomalous 
roughening is that it can be detected only by
determining the scaling of the structure factor.
Whenever such a scaling takes place in the 
problem under investigation
the new exponent $\alpha_s$ will only show up
when analyzing the scaling behaviour of
$S(k,t)$ and will not be detectable in 
either $W(L,t)$, $w(l,t)$ or $G(l,t)$.
In fact, as we have shown, the stationary
regime of a surface exhibiting this kind
of anomalous scaling will be characterized by
$W(L) \sim L^\alpha$ and 
$w(l,L) \sim \sqrt{G(l,L)} \sim l L^{\alpha-1}$,
however, the structure factor scales as
$S(k,L) \sim k^{-(2\alpha_s+1)}L^{2(\alpha-\alpha_s)}$
where the spectral roughness exponent 
$\alpha_s$ is a new and {\em independent} exponent.
We can summarize our analitycal results as follows 
\begin{equation}
\label{table}
\left\{
\begin{array}{lll}
{\rm if} & \alpha_s < 1 \Rightarrow \alpha_{loc}=\alpha_s & 
    \left\{ \begin{array}{l}
      \alpha_s=\alpha \Rightarrow {\rm Family-Vicsek}\\
      \alpha_s \neq \alpha \Rightarrow {\rm Intrinsic }\\
    \end{array}\right.\\
{\rm if} & \alpha_s > 1 \Rightarrow \alpha_{loc}=1 & 
    \left\{ \begin{array}{l}
       \alpha_s=\alpha \Rightarrow {\rm Super-rough}\\
       \alpha_s \neq \alpha \Rightarrow {\rm New\:\: class}\\
     \end{array}\right.\\
\end{array}\right.
\end{equation}

In the following we present simulations of a 
one-dimensional growth
model that is a nice example of the new dynamics.
We have performed numerical simulations of
the Sneppen model of self-organized depinning (model A)
\cite{sneppen}. 
We have found that this model exhibits anomalous
roughening the type described by Eq.(\ref{Anom-s}) for
$\alpha_s > 1$ and $\alpha_s \ne \alpha$.
In this model the height of the interface $h(i,t)$ is
taken to be an integer defined on a one-dimensional
discrete substrate $i=1, \cdots , L$. A random pinning
force $\eta (i,h)$ is associated with each lattice site 
$[i,h(i)]$. The quenched disorder $\eta (i,h)$ is 
uniformly distributed in $[0,1]$ and uncorrelated.
The growth algorithm is then as follows. 
At every time step $t$, the site 
$i_0$ with
the smallest pinning force is chosen and its height
$h(i_0,t)$ is updated $h(i_0,t+1) = h(i_0,t) + 1$
provided that the conditions $|h(i_0,t) - h(i_0 \pm 1,t)| < 2$
are satisfied. Periodic boundary conditions are assumed.
We have studied the behaviour of the model in systems
of different sizes from $L=2^6$ up to $L=2^{13}$. From calculations
of the saturated global width $W(L)$ for various
system sizes we find a global roughness exponent 
$\alpha = 1.000 \pm 0.005$ in agreement with previous 
simulations \cite{sneppen}. We have checked that 
the scaling of the global width
is given by Eq.(\ref{FV-globalwidth}) with a scaling function
like (\ref{FV-forf}). Also in agreement with previous work 
\cite{sneppen} we find that
the time exponent $\alpha/z = 0.95 \pm 0.05$. 
The local width $w(l,t)$ scales as
$w(l,t)=t^{\alpha/z} g(l/\xi)$ where the scaling function is
given by Eq.(\ref{a>1}), and also $\alpha_{loc} = 1$.

From these simulation results one could
conclude that the behaviour of the 
Sneppen growth model is rather
trivial and that the exponents $\alpha=\alpha_{loc}=z=1$
describe its scaling properties. Quite the opposite, 
this model exhibits no trivial features that can be
noticed when the structure factor is 
calculated. In Figure 1 we show our numerical results 
for the structure factor $S(k,t)$ 
in a system of size $L=2048$. 
Note that in Figure 1 the curves $S(k,t)$ for different 
times are shifted downwards reflecting 
that $\alpha < \alpha_s$. This contrasts with the case of
intrinsic anomalous roughening \cite{lopez97a,lopez97b}
where $\alpha_s=\alpha_{loc}$ and
$\alpha_{loc} \leq 1$. 
The slope of the 
continuous line is $-3.7$ and indicates that a new 
exponent $\alpha_s = 1.35$ enters the scaling. 
This can be better appreciated by the data collapse shown
in Figure 2, where one can observe 
that, instead of being constant, the scaling function $s(u)$
has a negative exponent $u^{-0.7}$ 
for $u \gg 1$. The exponents used for the data collapse are
$\alpha=1$, $z=1$ and the scaling function
obtained is in excellent agreement with 
Eq.(\ref{Anom-s}) and a spectral exponent $\alpha_s = 1.35 \pm 0.03 $. 

The interface in the Sneppen model A 
is formed by facets with constant slope $\pm 1$ \cite{sneppen}.
The value of the 
exponents $\alpha=\alpha_{loc}=1$ and $\alpha_s = 1.35$ 
is related to the faceted form of the interface at
saturation. It is easy to understand how the anomalous spectral 
roughness exponent appears due to the faceted form 
of the interface. For the simpler (and trivial) 
case of a faceted interface formed 
by a finite number of 
identical segments, $N$, of constant slope, $\pm m$, one can show
analitically that the global width
$W(L)\sim m^2 L^2/N^2$, and the height-height correlation
function $G(l) \sim l^2 m^2 - N m^2 l^3/L$, 
which leads to $\alpha=\alpha_{loc}=1$, while the spectrum
$S(k,L) \sim k^{-4} L^{-1}$ as $k \to 0$. A simple comparison
with the anomalous scaling form for the stationary spectrum
$S(k,L) \sim k^{-(2\alpha_{s}+1)} L^{2(\alpha-\alpha_{s})}$
leads to $\alpha_s = 1.5$. Actually, 
the facets occuring in the Sneppen model 
are not formed by identical segments, but rather 
follow a random distribution \cite{ramasco}, which leads to
a spectral exponent different from the trivial case.

%We have also found that these scaling properties 
%are shared by other models
%belonging to same universality class. In particular,
%we have studied the $1+1$ dimensional 
%Kardar-Parisi-Zhang equation
%\cite{kpz} with quenched noise and
%negative nonlinear term ($\lambda < 0$)
%which is believed 
%to describe the large scale dynamics of the Sneppen A
%model at the depinning transition \cite{jeong}.
%We have obtained similar anomalous scaling behaviour and 
%exponents for the pinned interfaces of the continuous equation
%just at the onset of depinning.
%We find $\alpha=\alpha_{loc} \simeq 1$ and a spectral 
%roughness exponent $\alpha_s \simeq 1.4$. 

In summary, we have presented a generic theory of
scaling for invariant surface growth. 
We have shown that 
the existence of power-law scaling of the correlation
functions (scale invariance) does not determine 
a unique form of the scaling functions involved.
This leads to the different dynamic scaling forms
recently observed 
in growth models 
\cite{krug,schro,lopez96,das1,das2,lopez97c,mario} 
and experiments 
\cite{yang,jef,fracture1,fracture2,bru}
exhibiting anomalous roughening.
In particular interface scale invariance does not 
necessarily imply Family-Vicsek dynamic scaling.
We have derived all the types of scaling (Family-Vicsek,
super-rough and intrinsic anomalous) from a unique
scaling ansatz, which is formulated in the Fourier space.
The different types of scaling are subclasses of our
generic scaling ansatz associated with bounds on
the values that the new spectral 
roughness exponent $\alpha_s$ may take. 
This generalization has allowed us to predict the existence of
a new kind of anomalous scaling with interesting
features. Simulations of a model for self-organized 
interface depinning have been shown to be in excellent
agreement with the new anomalous dynamics. 
It has recently been shown \cite{lopez99} 
that anomalous roughening  
stems from a non trivial dynamics of the mean local slopes
$\langle \overline{(\nabla h)^2} \rangle $. In contrast, the
new anomalous dynamics can be pinned down to growth models
in which the stationary state consits of faceted interfaces. 

The authors would like to thank R. Cuerno
for an earlier collaboration
that led to Ref.\cite{lopez97a,lopez97b}.
This work has been supported by the DGES of the
Spanish Government (project No. PB96-0378-C02-02). 
JJR is supported by
the Ministerio de Educaci\'on y Cultura (Spain).
JML is supported by a TMR Network of
the European Commission (contract number FMRXCT980183).

\begin{figure}
\centerline{
\epsfxsize=7.5cm
\epsfbox{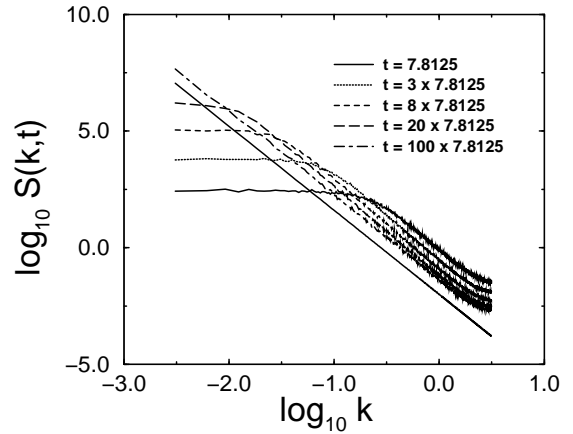}}
\caption{Structure factor of the Sneppen model for interface depinning
at different times. The continuous straight line is a guide to the eye 
and has a slope -3.7. 
Note the anomalous downwards shift of the curves 
for increasing times.}
\end{figure}

\begin{figure}
\centerline{
\epsfxsize=7.5cm
\epsfbox{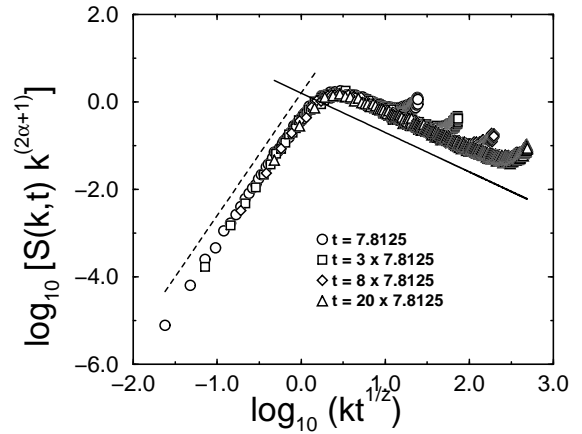}}
\caption{Data collapse of the graphs in Fig.\ 1.
The exponents used for the
collapse are $\alpha=1.0$ and $z=1.0$. The
straight lines have slopes $-0.7$ (solid) and $3.0$ (dashed) and are 
a guide to the eye. 
The scaling function is given by Eq.\ (7) with a 
spectral roughness exponent $\alpha_s = 1.35 \pm 0.03$.
The deviations from the scaling for large values of the 
argument $kt^{1/z}$ are due to the finite lattice spacing.}
\end{figure}

\end{multicols}
\end{document}